\documentclass[12pt]{article}
\usepackage{putex}
\usepackage{graphicx}
\usepackage{times}

\newcommand{\gws}{\gamma}
\newcommand{\deltax}{\delta x}

\begin{document}

\preprint{PUPT-2310}

\institution{PU}{Joseph Henry Laboratories, Princeton University, Princeton, NJ 08544, USA}

\title{Pointlike probes of superstring-theoretic superfluids}

\authors{Steven S. Gubser\footnote{e-mail: {\tt ssgubser@Princeton.EDU}} and
Amos Yarom\footnote{e-mail: {\tt ayarom@Princeton.EDU}}}

\abstract{In analogy with an experimental setup used in liquid helium, we use a pointlike probe to study superfluids which have a gravity dual.  In the gravity description, the probe is represented by a hanging string.  We demonstrate that there is a critical velocity below which the probe particle feels neither drag nor stochastic forces.  Above this critical velocity, there is power-law scaling for the drag force, and the stochastic forces are characterized by a finite, velocity-dependent temperature.  This temperature participates in two simple and general relations between the drag force and stochastic forces.  The formula we derive for the critical velocity indicates that the low-energy excitations are massless, and they demonstrate the power of stringy methods in describing strongly coupled superfluids.} 

\date{August 2009}

\maketitle

\section{Introduction}
Liquid Helium, at temperatures below $\sim 2.17 {}^{\circ} K$, is 
the paradigmatic example of a superfluid: a liquid with a broken global symmetry. Since its Helium atom constituents are strongly interacting it is difficult to obtain good theoretical control over the properties of the superfluid beyond the hydrodynamic approximation. This is exactly what the gauge-string duality provides: recent works \cite{Gubser:2008px,Hartnoll:2008vx,Basu:2008st,Herzog:2008he,Gubser:2009qm,Gubser:2009gp} offer theoretically tractable examples of a strongly coupled superfluid, holographically dual to superconducting black holes in five-dimensional anti-de Sitter space (AdS${}_5$) via the gauge-string duality \cite{Maldacena:1997re,Gubser:1998bc,Witten:1998qj}.

In this work we will use a holographic model of a superfluid to study its response to a heavy, pointlike probe.  Such an arrangement is similar to an experimental setup where a heavy ion is dragged through superfluid helium \cite{Allum}. Due to the low viscosity of the superfluid the ion moves without friction until it reaches the critical velocity for quasi-particle creation upon which it slows down. In superfluid helium, the quasi-particle excitations are rotons or vortices, depending on the details of the experimental setup.  We will show that in the holographic superfluids we study, there is also a critical velocity which can be determined from the bulk geometry but now the excitations associated with the onset of drag can be thought of as massless. We also study the stochastic force acting on the moving probe and extract from it the transverse and longitudinal mean momentum transfer.\footnote{The formulas for the mean momentum transfer have been obtained independently in \cite{HoyosBadajoz:2009pv} in a different context, and a significant simplification of the expression for $\kappa_L$ was carried out using techniques borrowed from that work.}

Holographic superfluids are different from liquid helium, being based on strongly coupled gauge theories.  Moreover, the infrared dynamics of the superfluids we study is characterized by emergent conformal symmetry.  The critical velocity that we compute from the low-temperature thermodynamics is the one that the emergent conformal group leaves invariant.  It defines an index of refraction of the holographic superfluid.  Thus we encounter a surprising confluence of ideas, weaving together superfluidity, emergent symmetry, black hole physics, string dynamics, and refractive signal propagation.

\section{The background geometry}

The gauge-string duality allows one to map a theory of gravity in an asymptotically five dimensional AdS space to a strongly coupled, large $N$, $SU(N)$ conformal field theory. Our starting point is the five dimensional bulk action $S = {1 \over 2\kappa_5^2} \int d^5 x \sqrt{-g} \, {\cal L}$ with
\begin{equation}
\label{GravTheory}
  {\cal L} = R - {1 \over 4} F_{\mu\nu}^2 - 
    {1 \over 2} \left[ (\partial_\mu \eta)^2 + \Sigma(\eta) 
      \left( \partial_\mu \theta - q A_\mu \right)^2 \right]  
    - V(\eta) \,.
\end{equation}
From here on, we will set $\kappa_5^2 = 1/2$.  In \cite{Gubser:2009qm} it was shown that \eqref{GravTheory} can be constructed from a truncation of IIB supergravity on a Sasaki-Einstein manifold.  (For a related example in M-theory, see \cite{Gauntlett:2009zw,Gauntlett:2009dn}.)  In this truncation, one finds that
$	\Sigma(\eta)  =  \sinh^2 \eta$,
$	V(\eta) = -{3 \over L^2} \cosh^2 {\eta \over 2} (5-\cosh\eta)$,
$	q=\sqrt{3}/L$
and the real scalars $\eta$ and $\theta$ describe a complex scalar whose target space is the Poincar\'e disk.  This complex scalar is charged under the $U(1)$ gauge field $A_\mu$ (where $F=dA$), so non-zero $\eta$ breaks the gauge symmetry. If a solution to the classical equations of motion of \eno{GravTheory} is found, then it can be lifted to a solution of type~IIB supergravity.
As explained in \cite{Gubser:2008px,Hartnoll:2008vx} the breaking of the $U(1)$ gauge-symmetry in the bulk corresponds to a formation of a condensate in the boundary theory.

We are interested in the bulk solution which corresponds to a zero temperature configuration where the charged scalar has no non-condensed component. In \cite{Gubser:2008wz} it was suggested that the zero-temperature limit of black hole solutions of theories similar to \eno{GravTheory} would be a domain wall with anti-de Sitter space on both sides: If we parameterize the line element, gauge field, and scalar field as
$  ds_M^2 = e^{2A(r)} \left( -h(r) dt^2 + d\vec{x}^2 \right) + dr^2 / h(r) $,
$  A_\mu dx^\mu = \Phi(r) dt $,
$\eta = \eta(r) $,
$\theta = 0 $
then the blackening function $h$ has positive slope everywhere, is approximately $1$ in the ultraviolet (UV) at large positive $r$ and asymptotes to a constant, which we denote $v_{\rm IR}^2$, in the infrared (IR) where $r$ is very negative.
The warp factor $A(r)$ also has positive slope everywhere and interpolates between two different linear functions of $r$, namely $r/L_{\rm IR}$ in the IR and $r/L$ in the UV.
Similarly, the scalar $\eta(r)$ rolls down from a local maximum of the potential $V$ in the ultraviolet to a local minimum in the infrared. Note that  $L_{\rm IR}$ and $\eta_{\rm IR}$ are related through $V(\eta_{\rm IR}) = -12/L_{\rm IR}^2$.
We will interpret this domain wall solution, in the dual gauge theory, as the ground state for a finite density of $U(1)$ charge, in which all the charge is carried by a superfluid condensate.

\begin{figure}
\begin{center}
\includegraphics[width=3 in]{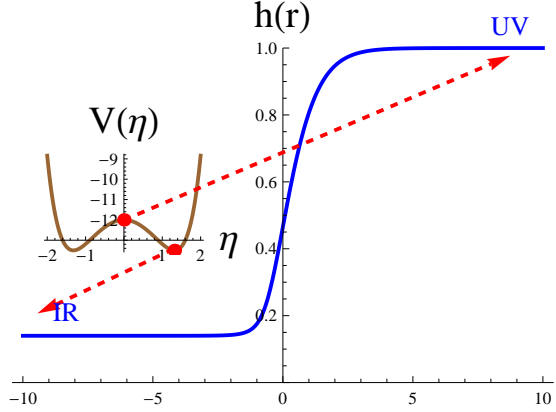}
\caption{\label{F:UVIRfig} A plot of $h(r)$ for the particular type IIB theory described in the text, at zero temperature. The scalar flows from an AdS${}_5$ region in the UV to an AdS${}_5$ region in the IR.}
\end{center}
\end{figure}

In the case of the IIB Lagrangian discussed above, $\eta_{\rm IR} = \cosh^{-1} 2$ and $\eta_{\rm UV} = 0$. An explicit domain wall solution for this theory was constructed in \cite{Gubser:2009gp}, where it was found that  $v_{\rm IR} \approx 0.373$.  This domain wall is a typical example of a zero temperature configuration with conformal invariance in the infrared as well as the ultraviolet. The result of \cite{Gubser:2009gp} has been reproduced in figure \ref{F:UVIRfig}.

\section{The drag force}
With the bulk description of a zero temperature superfluid in hand, we examine a heavy pointlike particle moving through the superfluid at constant velocity.  On the gravity side this is described by a trailing string, similar to the treatments of \cite{Herzog:2006gh,Gubser:2006bz}.  One end of the string is located at the asymptotically AdS${}_5$ boundary and specifies the location of the probe particle in the boundary theory. The other end of the string trails off into the deep IR. The momentum flowing down the string is equal in magnitude to the drag force acting on the boundary theory particle. Thus, to find the drag force one needs to look for an appropriate string configuration and compute the momentum flowing down the string. 

\begin{figure}
\begin{center}
\includegraphics[width=2.5 in]{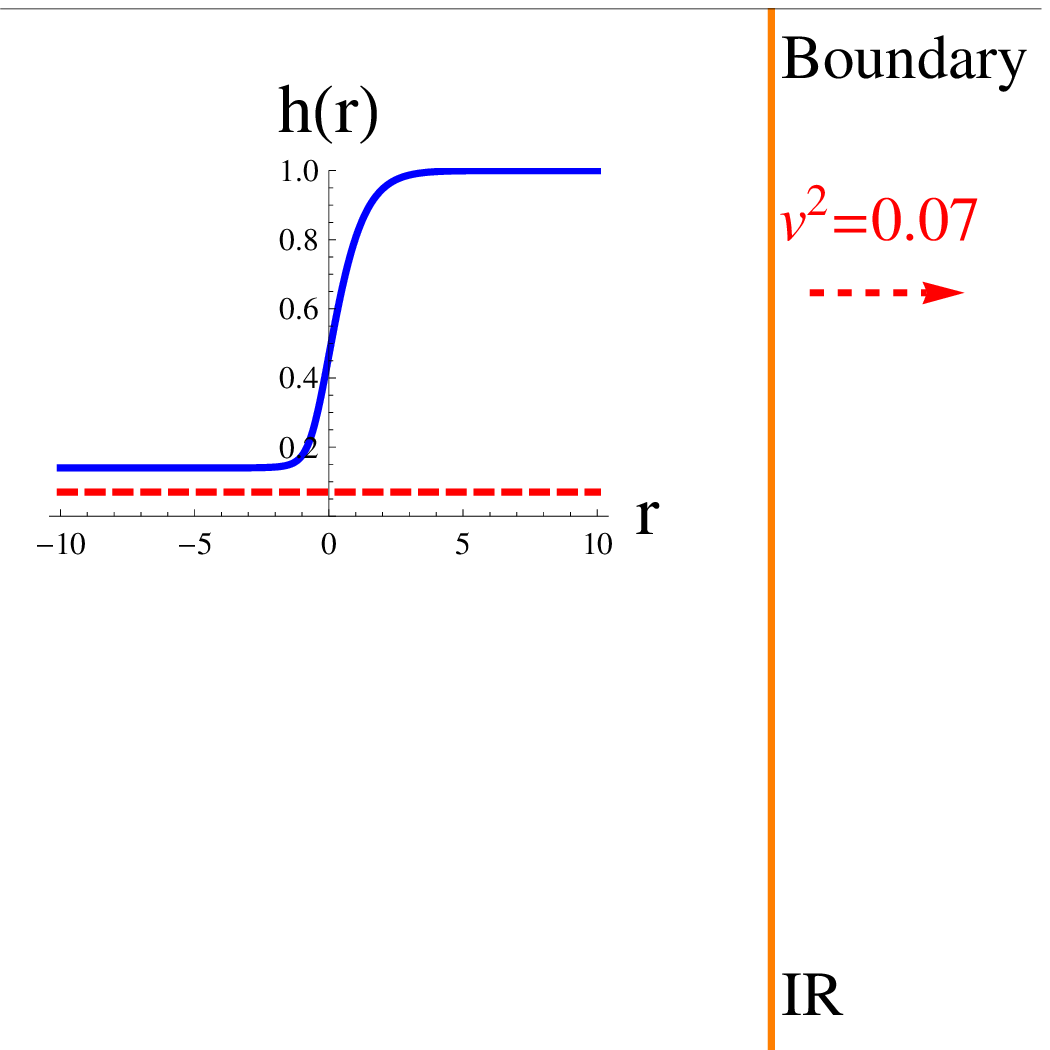}~~
\includegraphics[width=2.5 in]{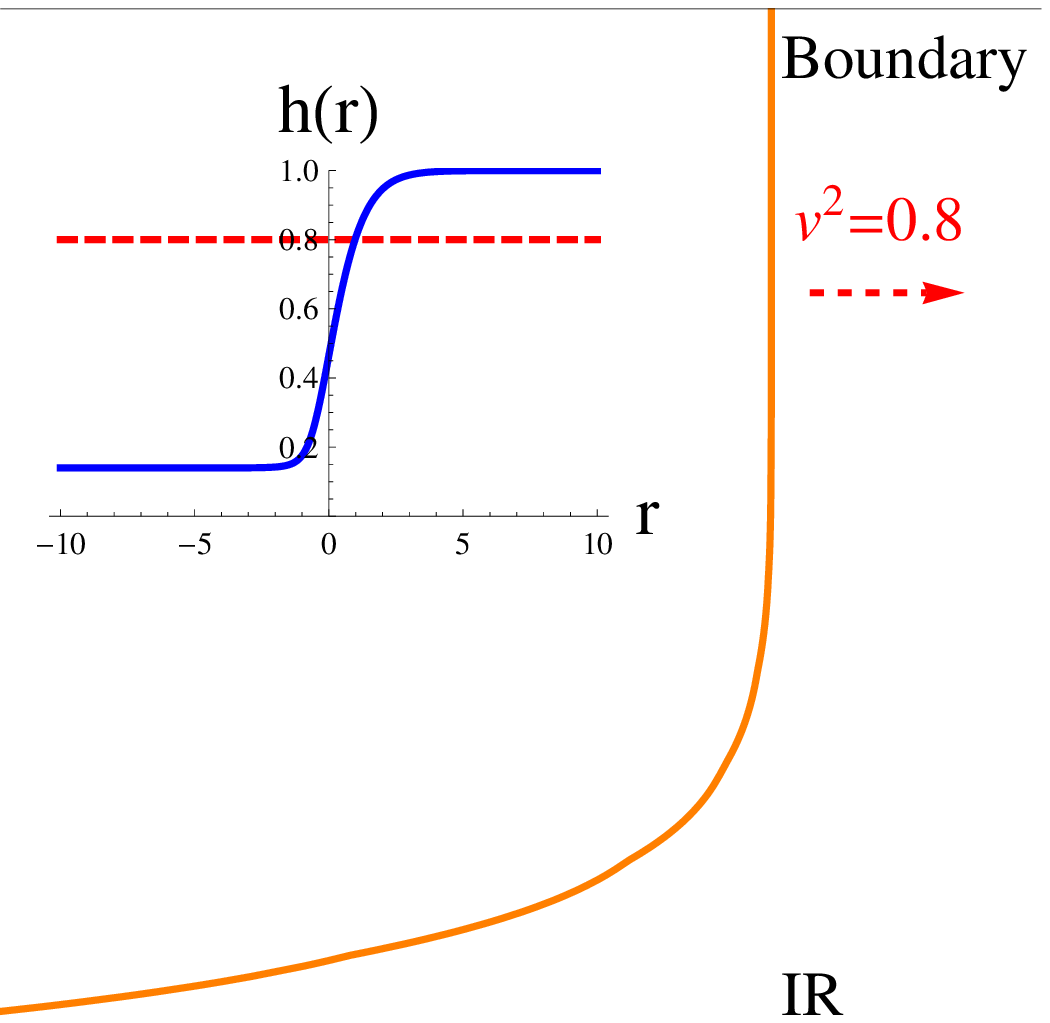}
\caption{\label{F:trailingfig} A plot of the trailing string with one endpoint on the boundary and the other in the IR region of AdS space for the IIB theory of \cite{Gubser:2009qm}. For $v<v_{\rm IR}$ there is no solution to $h(r_*)=v^2$ and the only possible configuration is a string hanging straight down and moving at constant velocity.}
\end{center}
\end{figure}

We will work with a string action of the form
\eqn{SNG}{
  S_{\rm string} = \int d\sigma d\tau \, {\cal L}_{\rm string} 
  \qquad\hbox{where}\qquad
  {\cal L}_{\rm string} = -{1 \over 2\pi\alpha'} 
  Q(\eta)
   \sqrt{-\det \partial_\alpha X^\mu \partial_\beta X^\nu g_{\mu\nu}} \,.
 }
Here $g_{\mu\nu}$ is the five-dimensional Einstein frame metric and $(\tau,\sigma)$ are worldsheet coordinates. In the particular case of the IIB action considered in \cite{Gubser:2009qm} one finds that $Q = \cosh {\eta \over 2}$. The procedure of extracting the drag force from the action is straightforward and here we merely quote the result:
\eqn{FoundFdrag}{
  F_{\rm drag} = -{e^{2A_*} Q_* \over 2\pi\alpha'} v \Theta(v-v_{\rm IR}) \,,
 }
where $A_*=A(r_*)$, $Q_*=Q(\eta(r_*))$, and so forth, and $r_*$ is determined implicitly by the equation $h(r_*)=v^2$. The result \eqref{FoundFdrag} (which was basically already understood in \cite{Herzog:2006se}) provides some interesting intuition about what warped geometry means: the string frame warp factor on the spatial dimensions $x^1$, $x^2$, and $x^3$ is $e^{2A} Q$, and, up to velocity-independent constants, this string frame warp factor evaluated at $r=r_*$ (when $r_*$ exists) is the coefficient of kinetic friction for a heavy pointlike probe. The surprising feature of \eqref{FoundFdrag} is the appearance of the step function $\Theta$ which vanishes when its argument is negative: Once the velocity of the probe is smaller than $v_{\rm IR}$ then $h_*=v^2$ has no solution, and the drag force vanishes. This is depicted in figure \ref{F:trailingfig}.
By Taylor expanding the drag force, the asymptotic behaviors of $F_{\rm drag}$ as $v$ approaches $1$ from below or $v_{\rm IR}$ from above can be written in terms of the infrared or ultraviolet parameters of the theory,
\begin{equation}\label{Ffast} 
  F_{\rm drag} = 
   \begin{cases}
    \displaystyle -\frac{\sqrt{\langle T_{00} \rangle L}}{2\sqrt{3} \pi \alpha'}\frac{v}{\sqrt{1-v^2}} + \ldots & v \to 1 \\[10pt]
   \displaystyle  -{Q (\eta_{\rm IR} ) v_{\rm IR} \over
   2 \pi\alpha'} \left(\frac{2}{b_h}\right)^{1/(\Delta_{\Phi}-4)} \left( {v  - v_{\rm IR}} \right)^{1/(\Delta_{\Phi}-4)} \Theta(v-v_{\rm IR}) & v \to v_{\rm IR}
   \end{cases}
\end{equation}
where $\langle T_{00} \rangle$ is the energy density of the zero temperature configuration in the boundary theory, $\Delta_{\Phi}$ is the dimension of the $U(1)$ current in the infrared defined so that a conserved current has dimension 3, and $b_h$ is an extra parameter which characterizes the deviation of $h$ from a constant value in the IR. For the type IIB theory of \cite{Gubser:2009qm}, $\Delta_{\Phi}=5$. The details of the derivation of \eqref{FoundFdrag} and \eqref{Ffast} can be found in appendix \ref{A:drag}.

The onset of the drag force in the holographic setup we've just described seems to be associated 
to a phenomenon analogous to roton emission and to Cherenkov radiation---the latter providing a slightly closer analogy, as we now explain.  Low-energy photons in a refractive medium comprise a free conformal field theory with a reduced speed of propagation relative to the speed of light in vacuum.
When an electrically charged particle travels through the medium faster than the speed of light, nearly free photons are emitted in a process called Cherenkov radiation.
What probably happens in our setup is that an operator which couples to the pointlike probe mediates strongly interacting ``radiation'' into the medium, and this radiation propagates at a velocity $v_{\rm IR}$.\footnote{We thank D. Son for discussions on this point.} 
Although this line of thought is plausible, it surprises us that the scaling $F_{\rm drag} \sim (v-v_{\rm IR})^a$ for $v$ slightly larger than $v_{\rm IR}$ is characterized by an exponent $a = (\Delta_\Phi-4)^{-1}$:  
The spectral measure of the two-point function of the operator that the probe couples to is the analog of phase space for photons emitted through the Cherenkov process. Naive phase space measure arguments applied to the case of weakly coupled quasi-particle emission suggest instead a scaling exponent $a \sim \Delta$, where $\Delta$ is the dimension of the operator responsible for the quasi-particle emission. An extension of these phase space arguments to strongly coupled theories, \`{a} la \cite{Georgi:2007ek}, still gives an $a \sim \Delta$ scaling where $\Delta$ would now be the dimension of the ``quasi-unparticles.''  Perhaps a better understanding of how ultraviolet cutoffs affect the strongly coupled radiation rate would resolve the quandary of the scaling of $a$; or perhaps there is an operator mediating the radiation whose anomalous dimension in the infrared is inversely proportional to $\Delta_\Phi$.  This issue certainly warrants further attention.

\section{Thermodynamics and the critical velocity}

In \cite{Carter:1995if} it was shown that a relativistic gas of particles in the grand canonical ensemble with dispersion relation $\omega = k c_q$ leads to a relation of the form $ c_q^2  = s T / (s T + \mu \rho)$ where $s$ is the entropy density, $T$ is the temperature, $\mu$ is the chemical potential, and $\rho$ is the charge density of the quasi-particles.  We now show, using arguments entirely on the gravity side, that a similar relation holds in the low-temperature limit for solutions of \eno{GravTheory}: 
\begin{equation}
\label{E:Gotvir}
	v_{\rm IR}^2 = \lim_{T \to 0} \frac{s T}{s T + \mu \rho_{\rm n}} \,,
\end{equation}
where $\rho_n$ is the density of $U(1)$ charge in the normal state, defined as in \cite{Herzog:2009md} by the thermodynamic relation
$
  dP = s \, dT + \rho_n \, d\mu - {\rho_s \over 2\mu} d\xi^2
$,
where $\xi_m = \partial_m \varphi$ and $\varphi$ is the Goldstone boson of the broken $U(1)$ symmetry. In our derivation, we will not require the assumption that there are low-energy quasi-particles that propagate with speed $v_{\rm IR}$.

To prove \eqref{E:Gotvir} we use the formalism developed in \cite{Herzog:2009md}. So far we considered the zero temperature solution of the theory. If the temperature is finite then the solution will be a black hole, by which we mean that the infrared limit of the theory is characterized in the bulk by a black hole horizon at $r=r_0$ where $h(r_0)=0$. The transition from a black hole solution to a domain wall solution occurs by the blackening function $h$ developing a ``double shelf'' structure.  That is, the horizon is pushed to the deep IR, before which $h$ is characterized by a region where it is approximately constant at a value approaching $v_{\rm IR}^2$. The double shelf structure was first discussed in \cite{Gubser:2008pf}.

In these positive but low temperature solutions, we consider vector perturbations of the stationary metric of the form $A_{x^1}=a(r)$ and $g_{tx^{1}} = e^{2 A(r)} g(r)$, and solve the linearized equations of motion for $g$ and $a$. The explicit form for the equation of motion for $g$ can be written as a total derivative by making use of the equations of motion for the background solution. After integrating it once it takes the form
$	e^{4 A}g'+e^{2 A}a \Phi' = Q_2 $
where $Q_2$ is an integration constant. By expanding the fields $a$ and $g$ near the boundary 
and using the relativistic two-fluid model for superfluids, introduced in \cite{Son:2000ht} and elaborated upon in \cite{Herzog:2008he,Yarom:2009uq,Herzog:2009md}, one can show that $Q_2 = s T u_{x^1} + (u_{x^1} - v_{x^1})\mu \rho_n$. Where $\vec{u}$ and $\vec{v}$ are the velocity of the normal component  and the superfluid velocity respectively. This leads to $Q_2 = s T u_{x^1} + (u_{x^1} - v_{x^1})\mu \rho_n$. Making use of Lorentz invariance, a similar analysis can be carried out for the equation of motion for $a$. One finds $h e^{2 A} (a \phi' - a'\phi) = h Q_2 - g Q_1 +Q_3 $ with $Q_3 = (v_x-u_x)\mu \rho_{\rm n}$. To obtain \eqref{E:Gotvir} we can evaluate the equation of motion for $a$ on the ``infrared shelf'' described earlier, and then take the zero temperature limit. The finer details of this analysis are left to appendix \ref{A:vir}.

Considering \eqref{E:Gotvir} together with the computation of \cite{Carter:1995if} and the expression for the drag force for $v$ near $v_{\rm IR}$, it is tempting to speculate that the low temperature thermal excitations are massless particles. However, one should keep in mind that our system is strongly coupled, so it's more likely that the critical velocity is associated with a strongly coupled ``soup'' of excitations whose collective behavior is in some ways similar to that of massless particles, but cannot be fully described in terms of quasi-particles.

\section{Stochastic forces}

Having determined the drag force acting on the probe particle and related the threshold velocity $v_{\rm IR}$ at which it becomes non-zero to thermodynamic quantities, we proceed to compute the stochastic forces on the probe.
In a Langevin formalism with white noise, one expresses the total force acting on the probe as
$  d\vec{p} / dt \equiv \vec{F}_{\rm tot} = \vec{F}_{\rm drag} + 
    \vec{F}_L(t) + \vec{F}_T(t) $, 
where $\vec{F}_L(t)$ is the stochastic force in the same direction as the motion of the probe particle while $\vec{F}_T(t)$ is orthogonal to it.  The expectation values of $\vec{F}_L$ and $\vec{F}_T$ vanish, while
$  \langle F_L^i(t_1) F_L^j(t_2) \rangle = \kappa_L \hat{p}^i \hat{p}^j
    \delta(t_1-t_2)  $ and
$  \langle F_T^i(t_1) F_T^j(t_2) \rangle = \kappa_T (\delta^{ij} - 
    \hat{p}^i \hat{p}^j) \delta(t_1-t_2) $,
where $\hat{p}^i$ is the unit vector in the direction of the momentum $\vec{p}$.  

In \cite{Casalderrey-Solana:2006rq,Gubser:2006nz,Casalderrey-Solana:2007qw} it was shown explicitly  that the stochastic force $\vec{F}_L + \vec{F}_T$ owes to the existence of a horizon on the string worldsheet: the endpoint of the string on the boundary is causally accessible to only that part of the string worldsheet with $r>r_*$, where $r_*$ is determined as before from $h_*=v^2$.  (If $v < v_{\rm IR}$, then there is no worldsheet horizon.  We will focus on the case $v_{\rm IR} < v < 1$ from now on.)  The metric on the string worldsheet above the horizon can be written in diagonal form
$  ds_{WS}^2 = \gws_{ab} d\sigma^a d\sigma^b = -e^{2A} (h-v^2) d\tau^2 + 
   \left( {1 \over h} + {e^{2A} h \xi'^2 \over h-v^2} \right) dr^2 
$,
where the worldsheet coordinates $(\tau,r)$ are specified by the indices $a,b$.  Note that $\tau=t$ only on the boundary.  The worldsheet Hawking temperature is given by
\begin{equation}\label{FoundTWS}
  T_{WS}    = \frac{e^{A_*}\sqrt{h'_*}}{4\pi} \left(h'_* + 4 v^2 A'_* + \frac{2v^2 Q'_*}{Q_*}\right)^{1/2} \,.
\end{equation}
Following arguments similar to the ones leading to \eno{Ffast}, one finds
\begin{equation}\label{TWSasymp}
T_{WS} = \begin{cases}
	\displaystyle\left( \frac{\langle T_{00} \rangle}{3 \pi^4 L}\right)^{1/4}  (1-v^2)^{1/4} & v \to 1 \\[10pt]
	\displaystyle\frac{ \sqrt{\Delta_{\Phi}-4} v_{\rm IR}^{3/2}}{\pi L_{\rm IR}} \left(\frac{2}{v_{\rm IR}b_h}\right)^{1/2(\Delta_{\Phi}-4)} (v-v_{\rm IR})^{1/2+1/2(\Delta_{\Phi}-4)}  & v \to v_{\rm IR} \,,
	\end{cases}
\end{equation}
provided $Q$ approaches a constant in the UV and IR. 
When $\Delta_{\phi}=5$ then $T_{WS}$ is linear in $v-v_{\rm IR}$ for $v$ close to $v_{\rm IR}$.

To study the stochastic forces further, one should consider perturbations of the string around the trailing string solution which lead to the drag force result \eqref{FoundFdrag}. Computations of this sort where carried out in \cite{Gubser:2006nz,Casalderrey-Solana:2007qw,Giecold:2009cg,Son:2009vu} and can be extended to general holographic backgrounds of the form we've discussed. The technique for carrying out this extension is to use the method of matched asymptotic expansions for approximate expressions for the fluctuations at small frequency and in the deep infrared. The result is
\begin{align}
\begin{split}
\label{GotKappa}
  \kappa_T& = -{2 F_{\rm drag} T_{WS} \over v} \\
  \kappa_L& = \kappa_T {\partial\log |F_{\rm drag}| \over \partial\log v} \,.
\end{split}
\end{align}
The details of this computation are left to appendix \ref{A:stochastic}.
We note that the form of $\kappa_T$ is interestingly similar to the Einstein relation
$  \kappa^{\rm Einstein}_L = -{2 F_{\rm drag} T / v} $,
which can be derived by demanding that stochastic evolution based on the total drag force $\vec{F}_{\rm  tot}$ equilibrates to a thermal distribution of temperature $T$, see for example \cite{Moore:2004tg}. But the differences between $\kappa_{T}$ and $\kappa^{\rm Einstein}_L$ are significant: besides constraining $\kappa_T$ rather than $\kappa_L$, the involvement of $T_{WS}$ in the expression for $\kappa_T$ makes it hard for us to see what the late-time equilibrium distribution of momenta would be when the probe mass is made finite.

\section*{Acknowledgments}
We thank C. Herzog, A. Nellore and D. Son for useful discussions. This work was supported in part by the Department of Energy under Grant No.\ DE-FG02-91ER40671 and by the NSF under award number PHY-0652782.

\begin{appendix}
\section{Details of the drag force computation}
\label{A:drag}
Our starting point is the string action given in \eqref{SNG}.
Consider the following ansatz for the string embedding.
 \eqn{TrailingString}{
 X^\mu(\tau,r) = \begin{pmatrix} \tau + \zeta(r) \\
  v  \tau +v \zeta(r) + \xi(r) \\
  0 \\
  0 \\
  r \end{pmatrix} \,,
 }
where we have set $\sigma=r$.  In the original treatments \cite{Herzog:2006gh,Gubser:2006bz}, the function $\zeta(r)$ was not introduced; indeed, the drag force computation is straightforward with $\zeta(r)$ set to $0$.  But, as pointed out in \cite{Gursoy:2009kk}, the choice
 \eqn{ZetaChoice}{
  \zeta(r) = -\int_r^\infty d\tilde{r} {v \xi'(\tilde{r})
    \over h(\tilde{r}) - v^2}
 }
makes the worldsheet metric diagonal, which in turn makes the treatment of fluctuations easier. In \eno{ZetaChoice} and hereafter, primes denote $d/dr$.  For the sake of consistency, we will work throughout with $\zeta(r)$ as given in \eno{ZetaChoice}.

Since $\xi'$ and not $\xi$ enters into ${\cal L}_{\rm string}$,
 \eqn{PiXiDef}{
  \pi_\xi \equiv {\partial {\cal L}_{\rm string} \over \partial \xi'}
 }
is constant when the equations of motion of the string are obeyed.  An analysis of the worldsheet current of spacetime energy-momentum analogous to the one in \cite{Gubser:2006bz} shows that $\pi_\xi$ is equal to the drag force $F_{\rm drag}$.  Solving \eno{PiXiDef} for $\xi'(r)$ leads to
 \eqn{XiPrime}{
  \xi'(r) = -{\pi_\xi \over h e^A} \sqrt{h-v^2 \over 
   h e^{4A} Q^2 / (2\pi\alpha')^2 - \pi_\xi^2} \,.
 }
Requiring that the solution is real everywhere and that only one string endpoint lies on the AdS boundary, we are lead to either $\pi_{\xi} = 0$, or
 \eqn{GotFdrag}{
  \pi_\xi = -{\sqrt{h(r_*)} e^{2A(r_*)} Q(\eta(r_*)) \over 2\pi\alpha'} \,,
 }
where $r_*$ is determined implicitly by the equation
 \eqn{GotV}{
  h(r_*) = v^2 \,.
 }
As in the main text, we will henceforth use a subscripted star to denote functions evaluated at $r=r_*$.  An important subtlety in \eqref{GotFdrag} is that if $v < v_{\rm IR}$, then there is no solution to \eno{GotV}.  In that case, the only solution to the classical equation of motion that extends arbitrarily far down into the bulk is the configuration with $\xi=\zeta = 0$.  If $v>v_{\rm IR}$ then the $\pi_{\xi} = 0$ solution is no longer physical because then the worldsheet metric is not Lorentzian for all $r$. Putting everything together, we find the drag force \eqref{FoundFdrag}.

The asymptotic behavior of $F_{\rm drag}$ as $v$ approaches $1$ is easy to understand based on the fact that the leading departure of the background geometry from pure AdS${}_5$ is
\begin{equation}
  h(r) = 1 - \frac{L \langle T_{00} \rangle}{3}  e^{-4A(r)} + \ldots \,, 
  \qquad
  A(r) = \frac{r}{L} + \ldots
\end{equation}
where $\langle T_{00} \rangle$ is the energy density of the dual field theory configuration. Using this expansion in \eqref{FoundFdrag} leads to \eqref{Ffast}.
Recall that we have set $\kappa_5^2 = 1/2$.

The behavior of $F_{\rm drag}$ as $v$ approaches $v_{\rm IR}$ depends on some detailed properties of the infrared geometry.  The key parameters are the infrared dimensions $\Delta_{\eta}$ and $\Delta_\Phi$ of the operators dual to $\eta$ and $A_\mu$; the bulk scalar and gauge field obtain an effective mass which can be computed by linearizing the equations of motion around the infrared fixed point.  For the particular geometry in \cite{Gubser:2009qm},\footnote{In \cite{Gubser:2008wz,Gubser:2008gr}, $\Delta_\Phi$ was chosen differently.  Our conventions here are such that $\Delta_\Phi=3$ when the photon is massless, which is when it corresponds to a conserved current $J_\mu$ in the dual field theory.}
 \eqn{DeltaValues}{
  \Delta_{\eta} = 2 + 2\sqrt{3} \qquad \Delta_\Phi = 5 \,.
 }
For general values of $\Delta_{\eta}$ and $\Delta_{\Phi}$ one finds the following solutions to the linearized equations in the IR:
 \eqn{InfraredExpansions}{
  A &= {r \over L_{\rm IR}} + b_A e^{\gamma_A r/L_{\rm IR}} + \ldots \cr
  h &= v_{\rm IR}^2 \left( 1 + b_h e^{\gamma_h r/L_{\rm IR}} + \ldots \right)  \cr
  \eta &= \eta_{\rm IR} + a_\eta e^{(\Delta_{\eta}-4) r/L_{\rm IR}} + \ldots  \cr
  \Phi &= a_\Phi e^{(\Delta_\Phi-3) r/L_{\rm IR}} + \ldots \,,
 }
where $\ldots$ denotes terms subleading to the ones shown.\footnote{There can also be an $r$-independent constant term in the expansion of $A$ which can be removed by gauge fixing $r$.} It is straightforward to show that
 \eqn{FoundGammaH}{
 	\gamma_h = 2(\Delta_\Phi-4) \,,\qquad
	\gamma_A = \min \left\{ 2 (\Delta_{\Phi}-4)\,,2 (\Delta_{\eta} - 4) \right\}\,.
 }
Also, $b_A$ and $b_h$ can be determined algebraically in terms of $a_\eta$, $a_\Phi$, $\Delta_{\eta}$, and $\Delta_\Phi$.  Using \eno{InfraredExpansions} one can express $v$ in terms of $b_h$,
 \eqn{ExpressV}{
  v  - v_{\rm IR} = \frac{1}{2} b_h e^{\gamma_h A_*} \,,
 }
to leading order in $v-v_{\rm IR}$.  Also at leading order, one has for $v \gsim v_{\rm IR}$
 \eqn{ExpressFdrag}{
  F_{\rm drag} = -{Q (\eta_{\rm IR} ) v_{\rm IR} \over 2\pi\alpha'}
    e^{2A_*} \,.
 }
Comparing \eno{ExpressV} and \eno{ExpressFdrag}, one arrives at \eqref{Ffast}.

\section{Some details on obtaining the expression for $v_{\rm IR}^2$.}
\label{A:vir}
To prove \eqref{E:Gotvir} consider a positive but low temperature background as discussed in the main text, but with a vector perturbation of the form
\begin{equation}
	A_{x^1}=a(r) \qquad\qquad g_{tx^{1}} = e^{2 A(r)} g(r)\,.
\end{equation}
The equation of motion for $g$ and $a$ is a set of two coupled second order differential equations.
As discussed in \cite{Herzog:2009md}, one of the solutions to this set of equations is a boost of the background metric
\begin{equation}
\label{E:Boost}
	a = - \phi u \qquad\qquad
	g = (h-1) u
\end{equation}
with $u$ an infinitesimal boost parameter in the $x^1$ direction (we use $\vec{x} = \left( x^1 \,, x^2 \,, x^3 \right)$). The explicit form for the equation of motion for $g$ can be written as a total derivative by making use of the equations of motion for the background. After integrating it once it takes the form
\begin{equation}
\label{E:Q2equation}
	e^{4 A}g'+e^{2 A}a \Phi' = Q_2
\end{equation}
where $Q_2$ is an integration constant. Expanding
\begin{equation}
	a = -\mu v_x +\frac{1}{2} \langle J_x \rangle  e^{-2 A} + \ldots \qquad
	\Phi = \mu + \frac{1}{2}\langle J_t \rangle e^{-2 A} + \ldots \qquad
	g = - \frac{1}{4} \langle T_{tx} \rangle e^{-4 A}+\ldots
\end{equation}
with $\langle J_{\mu} \rangle$ and $\langle T_{\mu\nu} \rangle$ the boundary theory charged current and stress tensor, we can obtain $Q_2$ in terms of boundary theory quantities. Using the relativistic two-fluid model for superfluids, introduced in \cite{Son:2000ht} and elaborated upon in \cite{Herzog:2008he,Yarom:2009uq,Herzog:2009md}, it is possible to express $\langle J_{\mu} \rangle$ and $\langle T_{\mu\nu} \rangle$ in terms of the hydrodynamic quantities: $s$, $T$, $\mu$, $\rho_{\rm n}$, the velocity of the normal component $\vec{u}$ and the superfluid velocity $\vec{v}$. This leads to $Q_2 = s T u_{x^1} + (u_{x^1} - v_{x^1})\mu \rho_n$. For the particular boosted solution \eqref{E:Boost} where $u_{x^1} = v_{x^1} \equiv u$, \eqref{E:Q2equation} reduces to
\begin{equation}
\label{E:Q1equation}
	e^{4A}h'-e^{2A}\Phi \Phi' = Q_1
\end{equation}
with $Q_1 = s T$.  The identity \eqref{E:Q1equation} can also be obtained through a Noether charge argument.\footnote{We thank A.~Nellore for explaining the Noether charge argument to us.} Using \eqref{E:Q2equation} and \eqref{E:Q1equation} the equation of motion for $a$ may also be integrated:
\begin{equation}
\label{E:Q3equation}
	h e^{2 A} (a \phi' - a'\phi) = h Q_2 - g Q_1 +Q_3
\end{equation}
where $Q_3 = (v_x-u_x)\mu \rho_{\rm n}$ is another integration constant which has been evaluated by expanding \eqref{E:Q3equation} near the AdS${}_5$ boundary. 

Next we evaluate \eqref{E:Q3equation} in the deep IR. There, 
\begin{equation}
\label{E:ShelfValue}
	\phi = \phi' = 0\,, \quad
	h = v_{\rm IR}^2\,, \quad
	g = (v_{\rm IR}^2 - 1)v_x\,.
\end{equation} 
The first two expressions follow from \eqref{InfraredExpansions} and the last expression follows from solving \eqref{E:Q3equation} and \eqref{E:Q2equation} with $Q_1=Q_2=Q_3=0$ and realizing that at zero temperature there is no normal component so that the only possible flow is given by the boosted solution \eqref{E:Boost} with the velocity of the superfluid component being the boost parameter.  As discussed in the main text, at low but non-vanishing temperatures one finds that $\phi$, $h$ and $g$ approach the values \eqref{E:ShelfValue} in the deep IR, before $h$ vanishes at the black hole horizon.
Thus, evaluating \eqref{E:Q3equation} for a positive but low temperature solution in the region where \eqref{E:ShelfValue} is valid, we obtain \eqref{E:Gotvir}.

\section{The stochastic force}
\label{A:stochastic}
To study the stochastic forces and how they lead to \eqref{GotKappa}, we consider perturbations of the string around the trailing string shape:
 \eqn{TrailingStringPerturbed}{
 X^\mu(\tau,r) = \begin{pmatrix} \tau + \zeta(r) \\
  v  \tau +v \zeta(r)  + \xi(r) + \deltax^1(\tau,r) \\
  \deltax^2(\tau,r) \\
  \deltax^3(\tau,r) \\
  r \end{pmatrix} \,.
 }
Plugging this ansatz into the action \eno{SNG} and expanding to quadratic order in the $\deltax^i$ leads to
 \eqn{LstringExpand}{
  {\cal L}_{\rm string} &= -{Q e^A \over 2\pi\alpha' \sqrt{h}}
   \sqrt{h-v^2+ e^{2A} h^2 \xi'^2} - 
   {K_L \over 2} (\partial_a \delta x^1)^2 - 
   \sum_{i=2,3} {K_T \over 2} (\partial_a \delta x^i)^2 + 
    {\cal O}(\deltax^3) \,,
 }
up to total derivative terms, where by $(\partial_a \deltax)^2$ we mean $\gamma^{ab} \, \partial_a \delta x \, \partial_b \delta x$, and 
 \eqn{GTGL}{
  K_L \equiv -{e^{2A} \over 2\pi\alpha'}
   {\sqrt{h_*} \over h} {Q_* \over \xi'}
  \qquad\qquad
  K_T \equiv -{e^{6A-2A_*} \over 2\pi\alpha'}
   {h \over \sqrt{h_*}} {Q^2 \over Q_*} \xi' \,.
 }
Note that $K_L$ and $K_T$ are both finite and smooth at the worldsheet horizon. In what follows we would like to treat the transverse and longitudinal directions simultaneously, so we will use a subscript $M=L,T$ where needed. The directions $x^2$ and $x^3$ are associated with $M=T$ while the $x^1$ direction is associated with $M=L$. 

Because neither $K_M$ nor $\gws^{ab}$ depends on $\tau$, we may express the general solution to the linearized equations of motion following from \eno{LstringExpand} as
 \eqn{ExpressGeneral}{
  \deltax^j(\tau,r) = \int_{-\infty}^\infty {d\omega \over 2\pi}
    \phi^j(\omega) e^{-i\omega\tau} \psi_M(\omega,r) \,,
 }
where by assumption $\psi_M(\omega,r) \to 1$ as $r \to \infty$ (the boundary of AdS${}_5$).  At the string worldsheet we require that $\psi_M$ has only infalling modes. The equation satisfied by $\psi_M(\omega,r)$ is
 \eqn{chiMeom}{
  \left( \partial_r K_M \gws^{rr} \partial_r - 
    K_M \gws^{\tau\tau} \omega^2 \right) \psi_M = 0 \,.
 }
This differential equation can be solved approximately for small $\omega$ using the method of matched asymptotic expansions. Briefly, the method is to match the far-field solution 
 \eqn{ChiFar}{
  \psi_M^{\rm far}(\omega,r) = 1 + C_{\rm far}(\omega) \int_r^\infty 
    {d\tilde{r} \over K_M(\tilde{r}) \gws^{rr}(\tilde{r})} \,,
 }
which solves \eno{chiMeom} to order $\mathcal{O}(\omega^2)$, with the near-horizon expansion
 \eqn{ChiNear}{
  \psi_M^{\rm near}(\omega,r) = 
    C_{\rm near}(\omega) (r-r_*)^{-i\omega/4\pi T_{WS}} \,.
 }
The integration constants $C_{\rm far}$ and $C_{\rm near}$ are $r$-independent, and in \eno{ChiNear} we have chosen the infalling solution, associated with the retarded two-point function.  Matching the near horizon behavior of \eno{ChiFar} to the small $\omega$ expansion of \eno{ChiNear} leads to
 \eqn{GotCfar}{
  C_{\rm far}(\omega) = {i\omega \over 4\pi T_{WS}} \left( K_{M} \,
   \partial_r \gws^{rr} \right) \big|_{r=r_*}+ {\cal O}(\omega^2) \,.
 }

Given a solution $\psi_M(\omega,r)$ to \eno{chiMeom}, there is an established way to extract both the two-point functions of the stochastic forces on the probe and the strength $\kappa_M$ of these forces in the low-frequency limit \cite{Gubser:2006nz,Casalderrey-Solana:2007qw,Giecold:2009cg,Son:2009vu}.  The symmetrized Wightman two-point function is
 \eqn{E:Greens}{
  G_M(\omega) = -\coth \left( \omega \over 2T_{WS} \right) 
    \Im G_M^{\rm ret}(\omega) = \coth \left( {\omega \over 2T_{WS}} \right)
   \lim_{r \to \infty} \Im\{ \psi_M^* K_M \gws^{rr} \partial_r \psi_M \} \,.
 }
Here $G_M^{\rm ret}(\omega)$ is the retarded two-point function.  Plugging \eno{ChiFar} into \eno{E:Greens} gives
 \eqn{E:GreensAgain}{
  G_M(\omega) = -\coth\left( {\omega \over 2T_{WS}} \right) 
     \Im C_{\rm far}(\omega) + {\cal O(\omega)} \,.
 }
The strength of the stochastic forces is now given by
 \eqn{GotKappaM}{
  \kappa_M = \lim_{\omega \to 0} G_M(\omega) = 
    -{1 \over 2\pi} \left(K_{M} \partial_r \gws^{rr}\right) \big|_{r=r_*} \,.
 }
Taking the $\omega \to 0$ limit means that we are approximating $G_M(t)$ by $\kappa_M \delta(t)$.  Obviously, this is valid only in the limit where we integrate the Langevin equation over a time longer than the characteristic timescale of $G_M(t)$, which is $1/T_{WS}$.

Using the expression for the worldsheet metric discussed in the main text and equations \eno{GTGL} \eno{FoundFdrag} and \eno{FoundTWS}, one obtains the expression for $\kappa_T$ in \eqref{GotKappa}. From \eno{GTGL} and \eno{GotKappaM} one can see immediately that
 \eqn{KappaLIntermediate}{
  \kappa_L = \left.\left({K_L \over K_T}\right) \right|_{r=r_*} \kappa_T = {\kappa_T \over 
    e^{2A_*} v^2 \xi^{\prime 2}_*} \,.
 }
To simplify this we start with \eno{XiPrime} and note that
 \eqn{LimXi}{
  \lim_{r \to r_*} {1 \over \xi'(r)} = -{v^2 e^{A_*} \over F_{\rm drag}}
    \sqrt{dF_{\rm drag}^2 / dr \over dv^2 / dr}
   = v e^{A_*} \sqrt{ {\partial\log |F_{\rm drag}| \over \partial\log v} } \,
 }
Which leads to $\kappa_L$ in \eqref{GotKappa}.

\end{appendix}

\newpage

\bibliographystyle{ssg}
\bibliography{hang}

\end{document}